# Large-Scale Query and XMatch, Entering the Parallel Zone

María A. Nieto-Santisteban,
Aniruddha R. Thakar
Alexander S. Szalay

*Johns Hopkins University*

Jim Gray

*Microsoft Research*

# Large-Scale Query and XMatch, Entering the Parallel Zone

María A. Nieto-Santisteban, Aniruddha R. Thakar, Alexander S. Szalay
*Johns Hopkins University*

Jim Gray
*Microsoft Research*

**Abstract**. Current and future astronomical surveys are producing catalogs with millions and billions of objects. On-line access to such big datasets for data mining and cross-correlation is usually as highly desired as unfeasible. Providing these capabilities is becoming critical for the Virtual Observatory framework. In this paper we present various performance tests that show how using Relational Database Management Systems (RDBMS) and a Zoning algorithm to partition and parallelize the computation, we can facilitate large-scale query and cross-match.

## 1. Introduction

The primary goal of the Virtual Observatory is to create the infrastructure necessary to make distributed digital archives accessible and interoperable in such a way that astronomers can maximize their potential for scientific discovery by querying and cross-matching multi-wavelength data between multiple archives on-the-fly. While small-area (few arcminutes or few thousand objects) searches are possible at present using tools like **Open SkyQuery**[1], large-scale requests involving all or a large fraction of the sky cannot be performed on demand due to system and network limitations.

## 2. Partitioning and Parallelization

The logical approach for speeding up access to data in large (multi-TB) datasets is to apply partitioning and parallelism within individual data services (SkyNodes). In this way, queries looking at different parts of the sky can be distributed among servers. Queries covering wide sky areas or full scans can be executed by different servers in parallel. Using the Zoning algorithm that we have developed, we can parallelize query and cross-match computations by distributing the data and workload among a cluster of database servers.

### 2.1. High-Speed Access Using Zones

The concept behind the Zoning algorithm is to map the celestial sphere into stripes of certain height called **Zones**. Each object at position *(ra, dec)* is assigned into a zone by using the fairly simple formula **ZoneID = floor ((dec + 90) / h)**, where *h* is the zone height.

Zoning the data has two main advantages when working with databases. First, the data and computation workload partition very easily by assigning different sets of zones to different servers and then executing the queries in parallel. When using Zoning to distribute the data, the height parameter, *h*, can be initially any value. After some tests, we chose *h = 4* arcminutes for our experiments which gives very good data distribution and computational performance. Second, when the database is relational, using zones helps to speed up neighborhood searches as explained in detail in (Gray et al. 2004). Once the objects have been assigned a ZoneID, we take advantage of the RDBMS high efficient indexing capabilities and build an index on ZoneID. This allows us to use pure relational algebra expressed in SQL to speed up simple queries like Cone Searches, or more sophisticated as those involved in finding clusters.

---

[1] http://www.openskyquery.net/

## 3. Test Case: SDSS, 2MASS, and USNO

We have used a vertical partition (subset of attributes) of the *Sloan Digital Sky Survey Data Release 3* (SDSS DR3), the *Two Micron All Sky Survey* (2MASS), and the *United States Naval Observatory B* (USNOB) catalogs to run large-scale access and cross-match queries. SDSS DR3 contains about 142 million objects covering a non-contiguous area of 5,282 squared degrees. In order to simplify our cross-match experiments, we used most of the 2MASS (28,445,694 objects) and USNO (117,698,363 objects) area that overlaps with SDSS DR3 (Figure 1-left). The next task was to distribute the workload in a homogeneous way. This step implied to zone the data and assign zone subsets to different servers in the cluster (Figure 1-right). At this point, the databases were replicated across the cluster.

The most efficient approach to query a single catalog is to distribute the workload homogenously so all nodes process the same amount of data. But cross-match queries require choosing a catalog leading the partitioning. What is the best partitioning choice is not always as simple as it might seem. For example, a reasonable choice would be to make the catalog with the smallest number of objects lead the partitioning. The assumption is that minimizing the size of the dataset works best because we reduce the number of operations on each server. This would be 2MASS in our test case. Looking at Figure 1, we can see that making 2MASS the leading partitioning catalog to do a cross-match with SDSS would be a terrible choice. While certain servers would be overloaded, others would basically remain idle. How to automatically decide what is the best distribution approach is not easy, if not impossible, without having precise information about the catalog densities and coverage footprints.

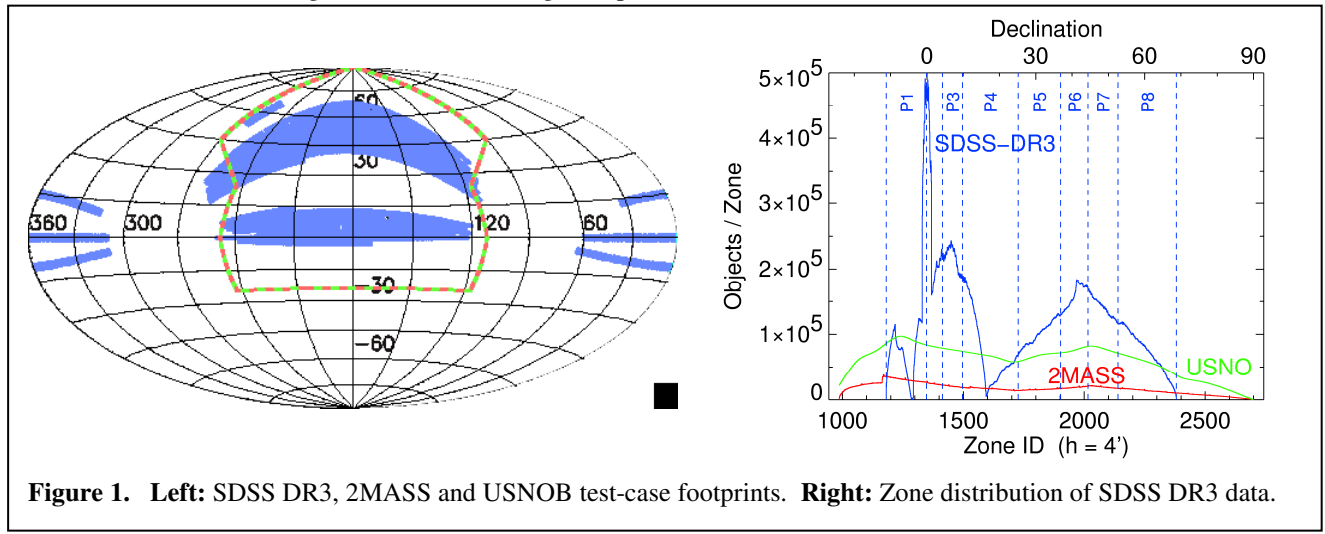

**Figure 1. Left:** SDSS DR3, 2MASS and USNOB test-case footprints. **Right:** Zone distribution of SDSS DR3 data.

## 4. Performance

### 4.1. Large-Scale Query

The query below requires a full table scan. It screens 142 million objects and retrieves only those meeting the magnitude filter. There is no index on the magnitude attribute.

```
SELECT p.objid, p.ModelMagR
INTO Results
FROM SDSSDR3:PhotoPrimary p
WHERE p.ModelMagR  BETWEEN 9.0 AND 10.0
```

According to Table 1, this query would take 11 minutes to be executed by a single server but 2 to 3 minutes when executed in parallel. While the latter is still far from being ideal, many users may be willing to wait for 3 minutes but very few will ever wait for 11 minutes in front of their browsers. It is important to note the big difference between elapsed and CPU time. This value shows that even though we can speed up the response time by incrementing the number of servers, the main reason for the "slow" response remains in the I/O bottleneck.

**Table 1. Full Scan Performance**

| Servers | Elapsed (s) | CPU (s) | I/O (MB) |     |
|---------|-------------|---------|----------|-----|
| 1       | 682         | 36      | 881      |     |
| 8       | 147         | 15      | 184      | MAX |
|         | 113         | 7       | 168      | AVG |

## 4.2. Large-Scale Cross-Match

We run the same cross-match procedures between SDSS DR3 and 2MASS using 1, 2, 4, and 8 servers to measure scalability. Figure 2 shows that scalability is certainly possible in terms of CPU time. This is directly related to a good workload distribution. On the center, the initial high slope for the one and two-server cases is due to I/O as right graph shows. However, the basically flat line plotting total elapsed time indicates that by using four or more servers we can as well speed up the cross-match computation linearly. It is worth noticing the small difference between **Total CPU** and **Total Elapse** that indicates good use of the CPU.

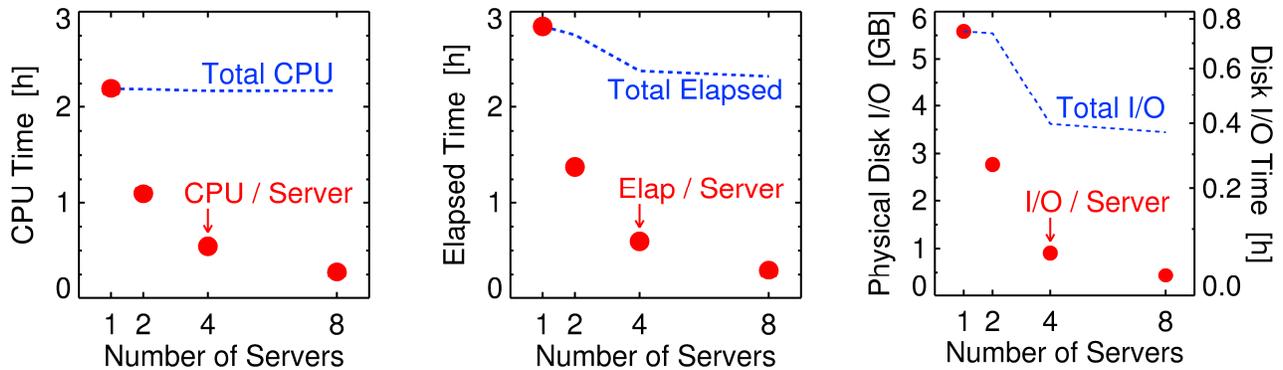

**Figure 2** SDSS vs 2MASS Cross-Match Performance.

## 5. Conclusions and Future Work

The tests and results presented here demonstrate that by zoning, partitioning and parallelizing the workload and using RDBMS technologies, we can reduce linearly the response time. However, the large-scale query and cross-match problem in the VO framework is still far from being completely resolved. In order to implement an interoperable system capable of managing the large-scale, the IVOA still needs to affirm protocols like the Asynchronous Activities which will provide a standard way to manage long jobs. The VOSpace/VOStore protocols and services are necessary so we can have locations where to put large query results, and means to share them with others. Authentication and Authorization mechanisms are essential to track who can access what and know what belong to whom. These protocols are under development and they will be implemented into our systems as they become recommendations.